\documentclass[12pt]{article}

\usepackage{graphicx}
\usepackage{graphics}
\usepackage{epsfig}
\usepackage{amssymb}
\usepackage{amsmath}
\usepackage{color}
\usepackage{textcomp}
\usepackage{epstopdf}

\usepackage[margin=2.5cm,footskip=0.6cm]{geometry}
\begin{document}

\begin{center}
{\Large{\bf Clustering and velocity distributions in granular gases cooling by solid friction}} \\
\ \\
\ \\
by \\
Prasenjit Das,$^1$ Sanjay Puri,$^1$ and Moshe Schwartz$^{2,3}$\\

$^1$School of Physical Sciences, Jawaharlal Nehru University, New Delhi 110067, India. \\
$^2$Beverly and Raymond Sackler School of Physics and Astronomy, Tel Aviv University, Ramat Aviv 69934, Israel.\\
$^3$Faculty of Engineering, Holon Institute of Technology, Golomb 52 Holon 5810201, Holon, Israel.\\
\end{center}

\begin{abstract}
We present large-scale molecular dynamics simulations to study the free evolution of granular gases. Initially, 
the density of particles is homogeneous and the velocity follows a Maxwell-Boltzmann (MB) distribution. The 
system cools down due to solid friction between the granular particles. The density remains homogeneous, and 
the velocity distribution remains MB at early times, while the kinetic energy of the system decays with time. 
However, fluctuations in the density and velocity fields grow, and the system evolves via formation of clusters 
in the density field and the local ordering of velocity field, consistent with the onset of plug flow. This is 
accompanied by a transition of the velocity distribution function from MB to non-MB behavior. We used equal-time correlation functions and structure factors of the density and velocity fields to study the morphology of 
clustering. From the correlation functions, we obtain the cluster size, $L$, as a function of time, $t$. We 
show that it exhibits power law growth with $L(t)\sim t^{1/3}$.
\end{abstract}

\newpage
\section{Introduction}
Granular materials consist of assemblies of particles with sizes ranging from 10$\mu m$ to 1$cm$~\cite{rev1}. This class of materials is probably the most important form of matter in the universe and the second most ubiquitous on Earth after water. The knowledge of its flow as a dense mass is relevant to phenomena such as mud, rock, and snow avalanches, transport of most technologically processed materials, transport of agricultural grains, etc,~\cite{rev2,rev3}. Low-density granular material appears in sand storms, smoke, and in interstellar sparse solid material. Although the flow of granular matter has similarities with the flow of ordinary fluids, there is a profound difference between the two and that is in the way energy is dissipated. In normal fluids mechanical energy is not really dissipated, it is just transferred from long into short scale disturbances of the flow that can be interpreted as heat. In granular matter, on the other hand, the interaction among the grains dissipates mechanical energy by storing it in intragrain degrees of freedom. The dynamical properties of granular systems have been studied experimentally by many authors. In this context, experimentalists have considered various standard geometries for agitating granular systems, e.g., horizontal and vertical vibration on a platform~\cite{1} pouring on an inclined plane and through a chute~\cite{2,3,4}, rotation in a drum~\cite{5,6,7}, etc. All of these experimental situations give rise to diverse examples of pattern formation, which have been of much research interest~\cite{8,9}.

The problem of granular gases, which is theoretically more accessible than the dense system, received 
considerable attention in the literature~\cite{10}. The leading theoretical approach was to concentrate on 
inelastic binary collisions, where the scattering of a pair of grains is characterized by a constant coefficient 
of restitution less than unity~\cite{11,12}. One of the most interesting problems studied within that approach 
is cooling and pattern formation. The system cools down due to the binary collisions, which are inelastic due to 
the incomplete restitution. In the early stages, it is in the homogeneous cooling state (HCS), with the uniform
density field. However, later in time, the density field becomes unstable to fluctuations and the system enters 
an inhomogeneous cooling state (ICS), where particle-rich clusters are formed and grow. The decay of temperature, which is the manifestation of the kinetic energy, has been well studied in the HCS~\cite{10} as well as  ICS~\cite{11}. Non-Maxwell-Boltzmann velocity distribution, e.g., power laws, stretched exponential, etc., have
been reported in various studies~\cite{13}. The complex pattern dynamics of density and velocity fields have been
studied by Puri \textit{et al.}, by invoking analogies from studies of phase ordering dynamics. The growth 
kinetics of the clusters in density and velocity fields have also been studied~\cite{14}.

In the case of dense flows, many particles rub against each other simultaneously and stay in prolonged contact. 
Thus, the concept of collision is not very useful. It seems, however, that the forces between touching grains 
are well understood: they can be decomposed in a force normal to the plane of contact and a solid friction force 
within that plane. Therefore, it may be expected that a continuum description can be achieved by coarse graining 
of the microscopic system~\cite{15,16}. In fact, a few continuous descriptions of the flow of dense granular 
matter, not inconsistent with the idea that solid friction is the mechanism of energy dissipation, had been 
published in the past~\cite{17,18}.

The high-density picture where solid friction is essential motivates our present study. In the case of granular gases cooling by inelastic binary collisions, it is well-known that the evolution of density, velocity, and granular temperature fields can be described by macroscopic hydrodynamic equations~\cite{32,33}. However, there are no corresponding hydrodynamic equations in the literature that can describe dense granular flow. An important and challenging issue in the physics of granular materials is to obtain a continuum description of dense granular flows. In this context, a major bottleneck has been a lack of proper understanding of the formation and evolution of plugs or clusters in dense granular flows. In this paper, we study the effect of the dissipation mechanism based on solid friction in granular gases. Although we believe that solid friction is the most significant dissipation mechanism in dense systems, we apply it here to gaseous granular matter, as the only mechanism of mechanical energy dissipation, to isolate this mechanism and study its effect on the cooling properties of such dilute systems. This will serve also as a test bed for future work on denser systems. Also, even in the present work, high-density regions appear due to clustering, to be discussed later, and in those regions solid friction should be indeed the dominant energy dissipation mechanism. We present results for velocity distributions, ordering dynamics of the density and velocity fields, and the growth dynamics of clusters.

This paper is organized as follows. In Sec. II, we describe our model and numerical results obtained therefrom. We focus on the evolution morphologies for the density and velocity fields in frictional granular gases, and their growth laws. In Sec. III, we conclude this paper with a summary and discussion.

\section{Frictional Cooling of Granular Gases}
We employ standard molecular dynamics (MD) techniques for our simulations, where all the particles are identical
with mass $m$. Any two particles with position vectors $\vec r_i$ and $\vec r_j$ interact via a two-body 
potential with a hard core and a thin shell repulsive potential. To be specific, we choose the potential to be of 
the following form
\begin{align}
 \label{pot}
 V(r) = \left\{
  \begin{array}{lr}
    \infty & : r < R_1\\
    V_0\frac{(r-R_2)^2}{(r-R_1)^2} & :  R_1 \le r < R_2\\
    0 &: r \ge R_2
  \end{array}
\right.
\end{align}
where r=$\mid\vec r_i - \vec r_j\mid$ is the separation between the two particles, $V_0$ is the amplitude of the 
potential, and $R_2 - R_1 < R_1$. Here, Eq.~(\ref{pot}) is to be taken only as a model of repulsive potential that
rises steeply from zero at the outer boundary of the shell to infinity at the hard core. The normal force applied 
by particle $i$ to particle $j$ is given by
\begin{eqnarray}
 \label{fn}
 \vec F_{ij}^n(r) = -\vec\triangledown V(r),
\end{eqnarray}
where the gradient is taken with respect to $r_j$. The corresponding solid friction force is given by
\begin{eqnarray}
 \label{ff}
 \vec F_{ij}^f(r) = \mu\mid \vec F_{ij}^n\mid \frac{\vec v_1 - \vec v_2}{\mid \vec v_1 - \vec v_2 \mid},
\end{eqnarray}
where $\vec v_i$ and $\vec v_j$ are the linear velocities of particles $i$ and $j$, respectively. Equation~(\ref{ff}) reduces to the well-known Coulomb's friction force when the thickness of the thin repulsive shell tends to zero, i.e., $R_2\rightarrow R_1$. In that limit, our model reduces to a hard sphere model where the velocity difference cannot have a normal component at the contact point. Thus, the frictional force becomes tangential to the normal force. For simplicity, we did not consider rotational motion of the grains. We use the following units for various relevant quantities: lengths are expressed in units of $R_1$, temperature in $V_0/k_B$, and time in $\sqrt{mR_1^2/V_0}$. For the sake of convenience and numerical stability, we set $R_1=1$, $R_2=1.1$, $V_0=10$, $k_B=1$, and $m=1$. The velocity Verlet algorithm~\cite{19} is implemented to update positions and velocities of the MD simulations. The integration time step is $\Delta t=0.0005$. The granular gas consists of $N=250000$ particles confined in a 2D box with periodic boundary conditions. Two area number densities $\sigma=0.20$ and $\sigma=0.30$ corresponding to area fraction $\phi\approx0.157$ and $\phi\approx0.236$, respectively, were considered. This means that the box sizes are $1118^2$ and $912^2$, respectively.

\begin{figure}
\centering
\includegraphics*[width=0.6\textwidth,height=0.6\textwidth]{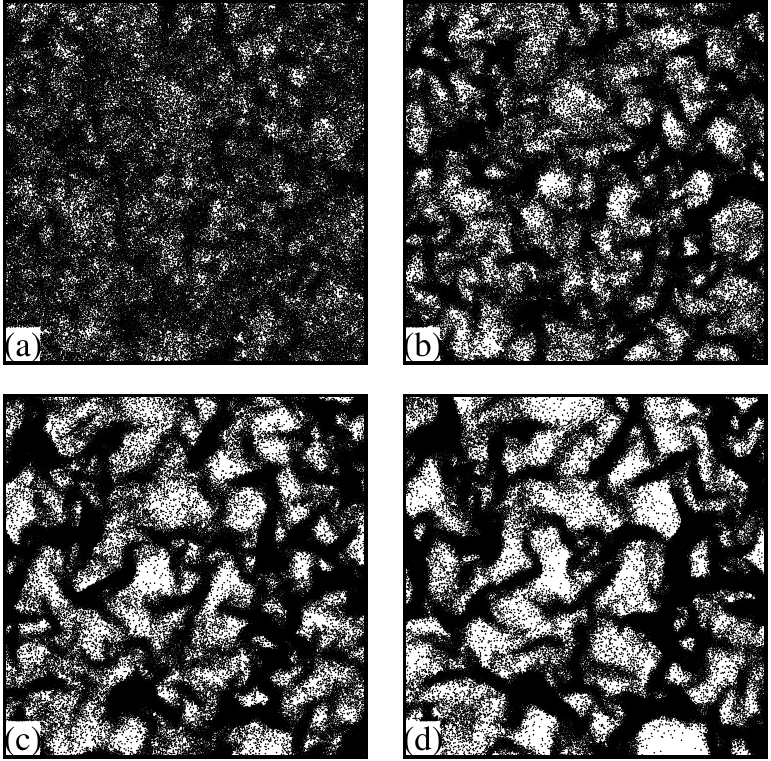}
\caption{\label{fn1} Evolution snapshots of the density field for an inelastic granular gas in $d=2$ at different
times: (a) $t=50$; (b) $t=150$; (c) $t=300$; (d) $t=500$. These pictures are obtained for a system with particle number $N=250000$, number density $\sigma=0.30$ (packing fraction $\phi\approx0.236$), and friction coefficient $\mu=0.10$. The size of the system is $912^2$. For clarity, we have shown only a $600^2$ corner. The density field is obtained by directly drawing a black point at the center of particle. Void spaces represent, thus, regions free of particles.}
\end{figure}

The system is initialized by randomly placing particles in the simulation box, such that there is no overlap 
between the cores of any two particles. All these particles have the same speed but the velocity vector points in 
random directions so that $\sum_{i=1}^N \vec v_i = 0$. The system is allowed to evolve until $t=50$ with $\mu = 0$, i.e., the elastic limit. The system is thus relaxed to a Maxwell-Boltzmann (MB) velocity distribution, which serves as the initial condition for our MD simulation of inelastic spheres with $\mu\ne0$.

Starting from the homogeneous initial condition in thermal equilibrium at $t=0$, the system starts dissipating 
its energy because of the frictional collisions among the particles for $\mu \ne 0$. Figure~\ref{fn1} shows the 
evolution snapshots of the density field for two-dimensional granular gas with $\mu = 0.10$. Details are given 
in the figure caption. 

At the early stage of evolution, the system remains roughly homogeneous. However, at the later stage, the 
formation of clusters is observed. This cluster formation can be explained as follows. Consider some fluctuations
in the homogeneous phase of the density. Particles in the high-density regions lose more energy than those in 
the low-density regions. Particles entering randomly from the adjacent low-density surrounding regions have finite 
probability to be trapped within the high-density region. That probability increases with the number of particles
in the region, thus increasing the density fluctuations. This effect is amplified by the fact that as time goes on, the system becomes sluggish due to the dissipation of energy and consequently the probability for a particle to be trapped in the high-density region increases.

\begin{figure}
\centering
\includegraphics*[width=0.6\textwidth,height=0.6\textwidth]{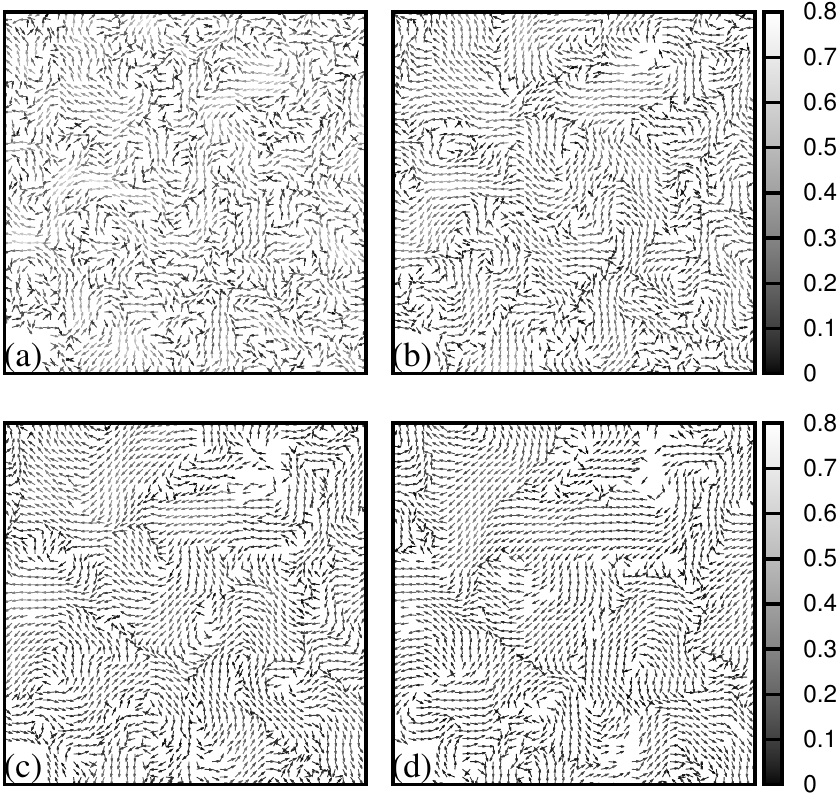}
\caption{\label{fn2} Evolution snapshots of the coarse-grained velocity field at different times: (a) $t=50$; (b) $t=150$; (c) $t=300$; (d) $t=500$. For the sake of clarity, we have shown only a $48^2$ corner of the $128^2$ box. }
\end{figure}

Figure~\ref{fn2} shows the coarse-grained evolution snapshots of the velocity field for a two-dimensional granular 
gas with $\mu=0.10$. The system is divided into squares of area $7.125^2$ and the average velocity is calculated
for each square. The direction of the average velocity is plotted as an arrow starting at the center of each 
square. The size of the average velocity is described by a shade of gray scale. The darker the shade of gray, the
lower the velocity. Void spaces represent, as in the density snapshots, regions free of particles.

At the early stage of evolution, the velocity field remains random. Correlations develop in the velocity field 
at a later time because solid friction between touching (shell overlap) particles causes velocity matching. Thus,
local ordering in the velocity field is observed, and the evolution of the velocity field is characterized by
the emergence and diffusive coarsening of vortices. Of course, the overall momentum is conserved, and this must 
be reflected in the ordered state also.

\begin{figure}
\centering
\includegraphics*[width=0.5\textwidth,height=0.4\textwidth]{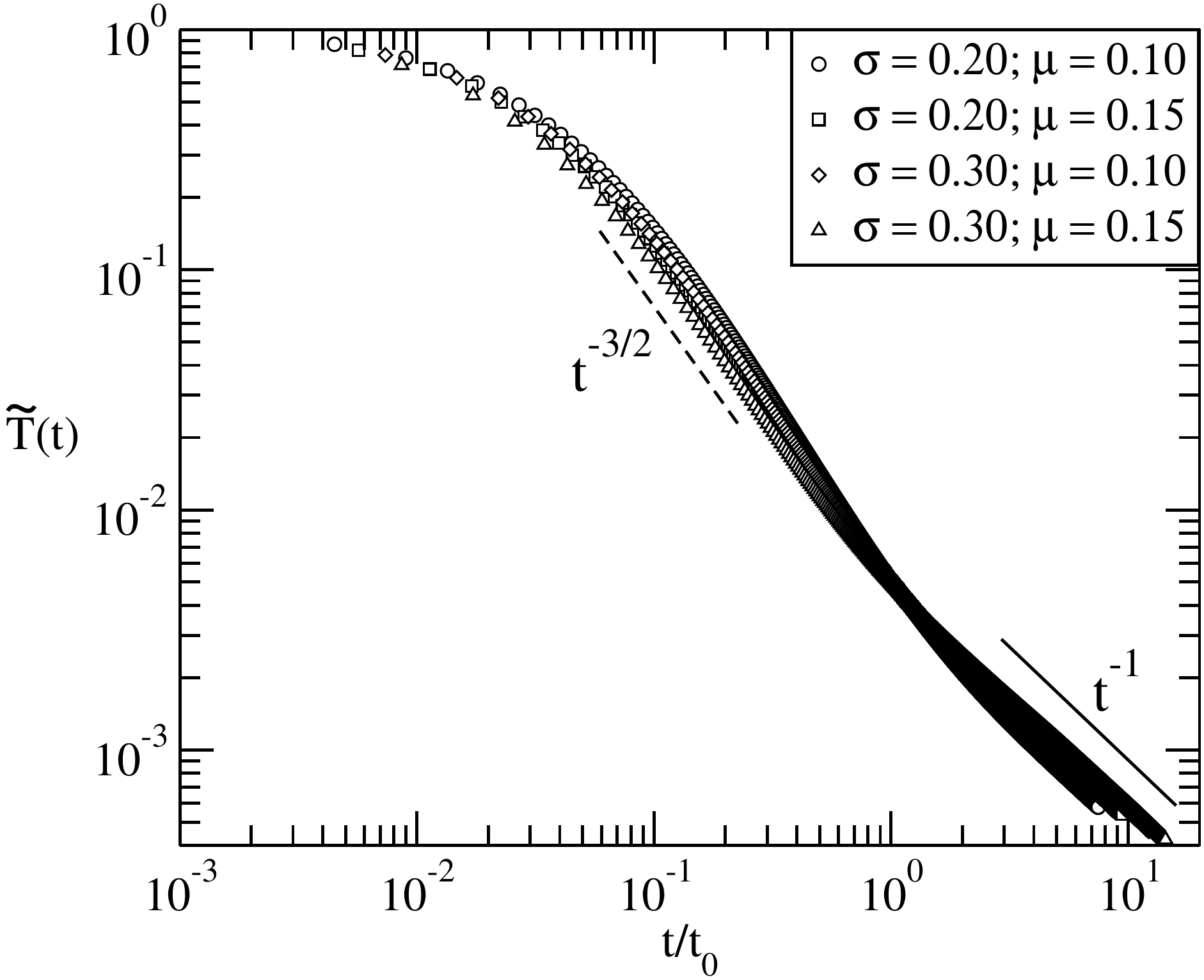}
\caption{\label{fn3} The scaling form of the temperature decay as a function of scaled time $t/t_0$. The crossover time is defined in the text, and is larger for smaller $\sigma$ and $\mu$. At early times, temperature decays as $\tilde T(t)\sim \left(t/t_0\right)^{-3/2}$. As time increases the decay slows down, leading eventually to a decay of the form $\tilde T(t)\sim \left(t/t_0\right)^{-1}$. Dashed line and solid line, respectively, represent algebraic decays with exponent $3/2$ and $1$.}
\end{figure}

Following the qualitative study described above, we go into a more quantitative study starting with the free 
cooling of the inelastic granular gases with nonzero values of $\mu$. The granular temperature is defined as 
$T=\langle\vec v ^2\rangle/d$, where $\langle\vec v ^2\rangle$ is the mean-squared velocity of a grain and $d$ 
is the spatial dimension. We present in Fig.~\ref{fn3} the cooling of the system by giving the temperature of 
the system as a function of time,
\begin{eqnarray}
\label{temp}
 T(t)=T(0)\tilde T(t),
\end{eqnarray} 
where $T(0)$ is the initial granular temperature. We find that for given $\sigma$ and $\mu$, $\tilde T(t)\propto t^{-\alpha}$, where the exponent $\alpha=3/2$ at short times crosses over to $\alpha=1$ at longer times. We define the crossover time as the time when the local exponent $\alpha$ equals $1.25$. The crossover time, $t_0$, is larger for smaller $\sigma$ and $\mu$. In Fig.~\ref{fn3}, we present $\tilde T(t)$ as a function of $t/t_0$ for four ($\sigma$, $\mu$) combinations. The early-stage cooling [$\tilde T(t)\propto t^{-3/2}$] corresponds to the \textit{homogeneous cooling state} (HCS) and is the counterpart of Haff's law for inelastic hard spheres. For the latter system, Haff's cooling law is as follows: $\tilde T(t)\propto t^{-1}$~\cite{10,13,14}.

The late-stage cooling [$\tilde T(t)\propto t^{-1}$] arises for the \textit{inhomogeneous cooling state} (ICS), where the density field has undergone the clustering instability. A similar late-stage cooling has been observed by Nie \textit{et al.}~\cite{29} for the inelastic hard sphere system. To avoid inelastic collapse, Nie \textit{et al.} modeled the collisions as elastic when the relative collision velocity is less than a critical value. In the late stages of evolution in our present model, plug formation has been observed. Since there is no relative velocity among the particles in a given plug, particles do not lose energy as in the case of elastic collisions. A similar late-stage cooling has also been observed by Baldassarri \textit{et al.}~\cite{30} in the context of ordering in a lattice granular fluid. 

\begin{figure}
\centering
\includegraphics*[width=0.8\textwidth]{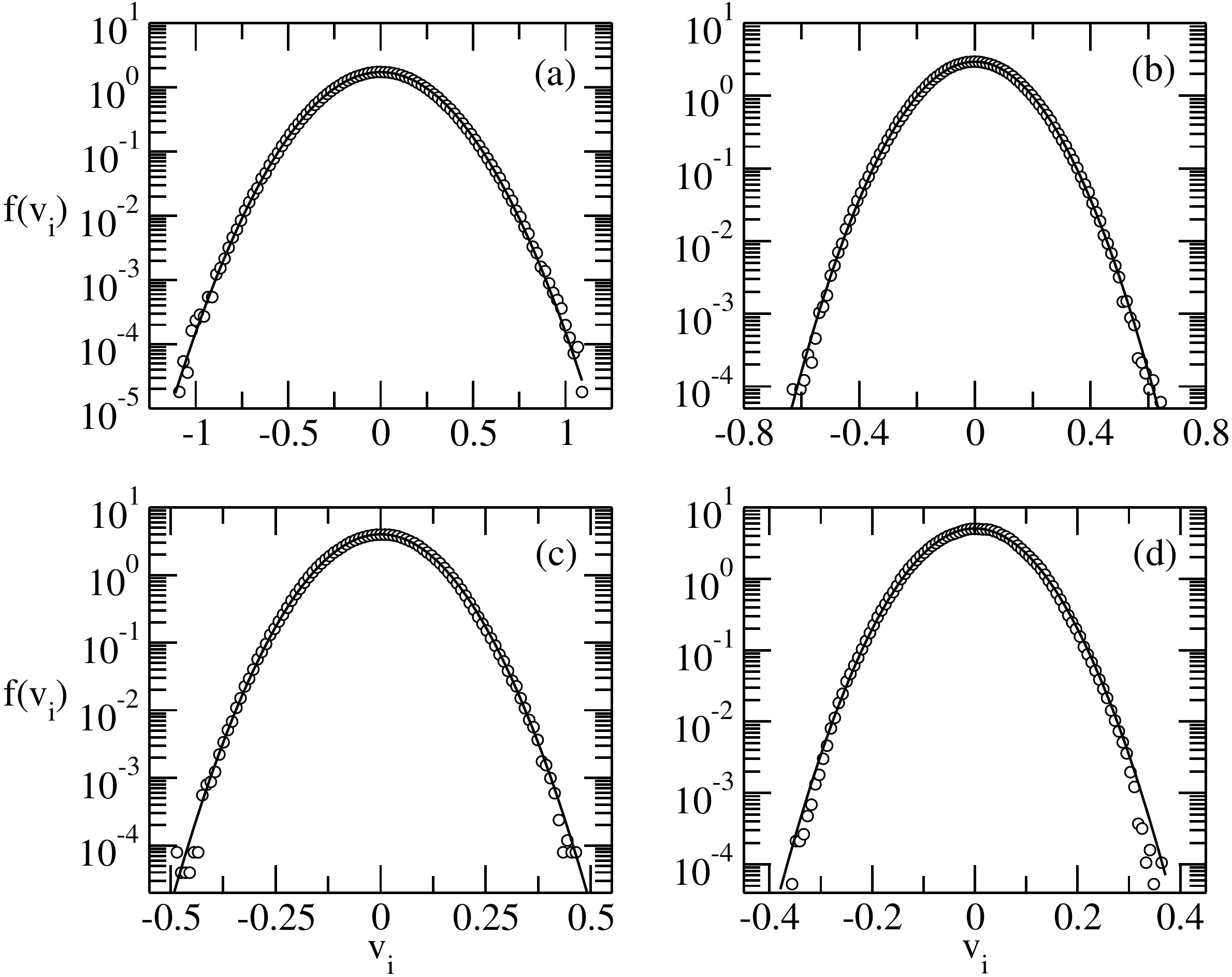}
\caption{\label{fnv} Normalized velocity distribution function $f(v_i)$ plotted on a semi-log scale for (a) $t=50$, (b) $t=150$, (c) $t=300$, and (d) $t=500$. Data is shown for $\mu=0.10$ and $\sigma=0.30$ at different times. The open circles represent the data obtained from the numerical simulations and averaged over ten independent runs. The solid line represents the $P_\text{MB}$ at that time (with the corresponding time dependent temperature). The statistical data deviates from $P_\text{MB}$ at later times, particularly in the tail region.}
\end{figure}

After establishing the time dependence of the temperature, we studied the time evolution of the velocity 
distribution function. The natural framework to study velocity distributions for the elastic granular gas is the
MB equation. In the elastic case with $\mu = 0$, an arbitrary initial velocity distribution rapidly evolves 
(after a certain time) to the MB distribution:
\begin{eqnarray}
 \label{mbd}
 P_{\text{MB}}(v_i) = \sqrt{\frac{m}{2\pi k_B T}}\exp\left(-\frac{mv_i^2}{2k_BT}\right),
\end{eqnarray}
where $v_i=(v_x,v_y)$ are the components of the velocity $\vec v$. For inelastic granular gases with $\mu\ne0$,
because of cooling the velocity distribution functions are time-dependent. It may be expected, at first sight, 
that the velocity distribution will depend on time only through the time-dependent temperature $T(t)$. The 
clustering phenomena discussed above suggests, however, that each particle may be viewed now as belonging effectively to one of a set of large super-particles, each with essentially a different mass. Thus, at short times, when clustering is not yet pronounced, a MB distribution with a time-dependent temperature may be expected, but once clustering sets in, the nature of the distribution is expected to change. Figure~\ref{fnv} shows the time evolution of the velocity distribution function for $\sigma=0.30$ and $\mu=0.10$. At early times, the velocities follow the MB distribution. However, at later times, deviations from the MB distribution are seen in the tail region, e.g., compare the data sets for $t=50$ and $t=500$. This is consistent with our earlier simulations of the freely-cooling inelastic hard-sphere gas~\cite{14}.

\begin{figure}
\centering
\includegraphics*[width=0.5\textwidth]{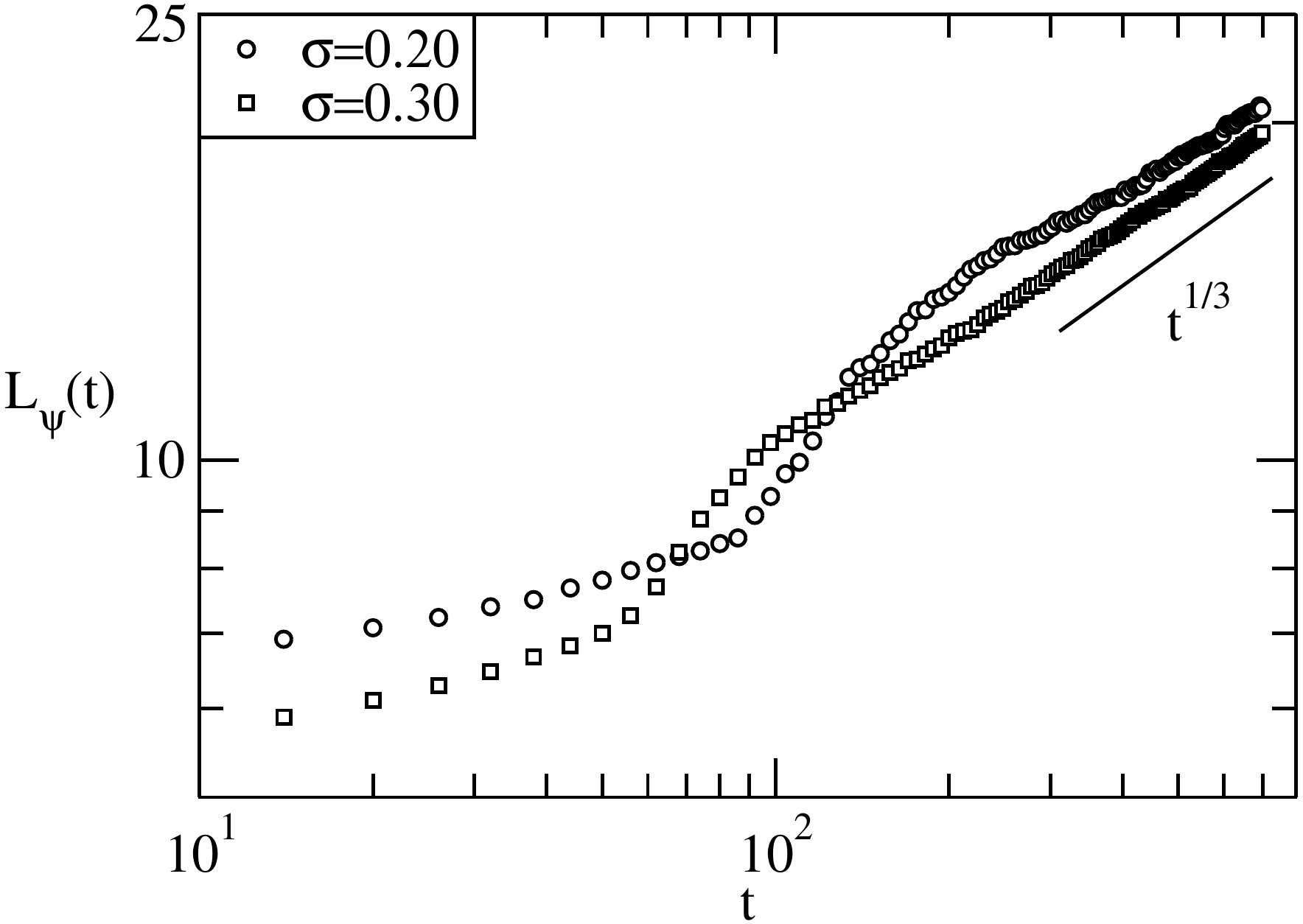}
\caption{\label{fnlp} Time-dependence of the correlation length $L_\psi(t)$ corresponding to the clustering
of the density field order-parameter, $\psi(\vec r, t)$ field. Clearly, in the late stage, $L_\psi(t)$ follows a
$t^{1/3}$ growth law. The solid line with exponent $\frac{1}{3}$ is shown, corresponds to diffusive growth of 
clustering.}
\end{figure}

The clustering of the density field $\sigma(\vec r, t)$ and velocity field $\vec v(\vec r, t)$ has been studied by invoking an analogy from phase-ordering systems with scalar and vector order parameters, respectively~\cite{14,31}. We studied the evolution morphologies of the $\sigma(\vec r, t)$ field and the $\vec v(\vec r, t)$ field by calculating equal-time correlation functions and structure factors~\cite{20,32,33}. For 
$\sigma=0.30$, the coarse-grained fields at a lattice point are obtained by calculating the average density and
the velocity within boxes of size $7.125^2$. We introduce the order parameter $\psi(\vec r, t)$  that attains the
values $+1$($-1$) where the local number density is more than (less than) the average number density ($\sigma_\text{av} \sim 0.30$, in this case), similar to a two-state Ising model~\cite{20}. This hardening of the order-parameter field is done to clearly extract the Porod tail in the structure factor~\cite{21,22}. The evolution of the $\sigma(\vec r, t)$ field is characterized by the order-parameter correlation function $C_{\psi\psi}(r, t)$, defined as
\begin{eqnarray}
 \label{denc}
 C_{\psi\psi}(r, t) = \left\langle\psi\left(\vec R, t\right)\psi\left(\vec R + \vec r, t\right)\right\rangle - \left\langle\psi\left(\vec R, t\right)\right\rangle\left\langle\psi\left(\vec R + \vec r, t\right)\right\rangle,
\end{eqnarray}
where the angular brackets represent the averaging over different initial conditions. Consider first the 
typical length scale, $L_\psi(t)$ characterizing the clustering. It is defined as the distance over which 
$C_{\psi\psi}(r, t)$ decays from $1$(at $r=0$) to $0.25$. The variation of $L_\psi(t)$ with $t$ is shown in
Fig.~\ref{fnlp} and shows a power-law behavior according to $L_\psi(t)\sim t^{1/3}$. This behavior corresponds
to the diffusive growth of particle rich clusters. (Note that this is exactly the behavior predicted for the 
growth of plug regions in granular flow \cite{16}.)

\begin{figure}
\centering
\includegraphics*[width=0.7\textwidth]{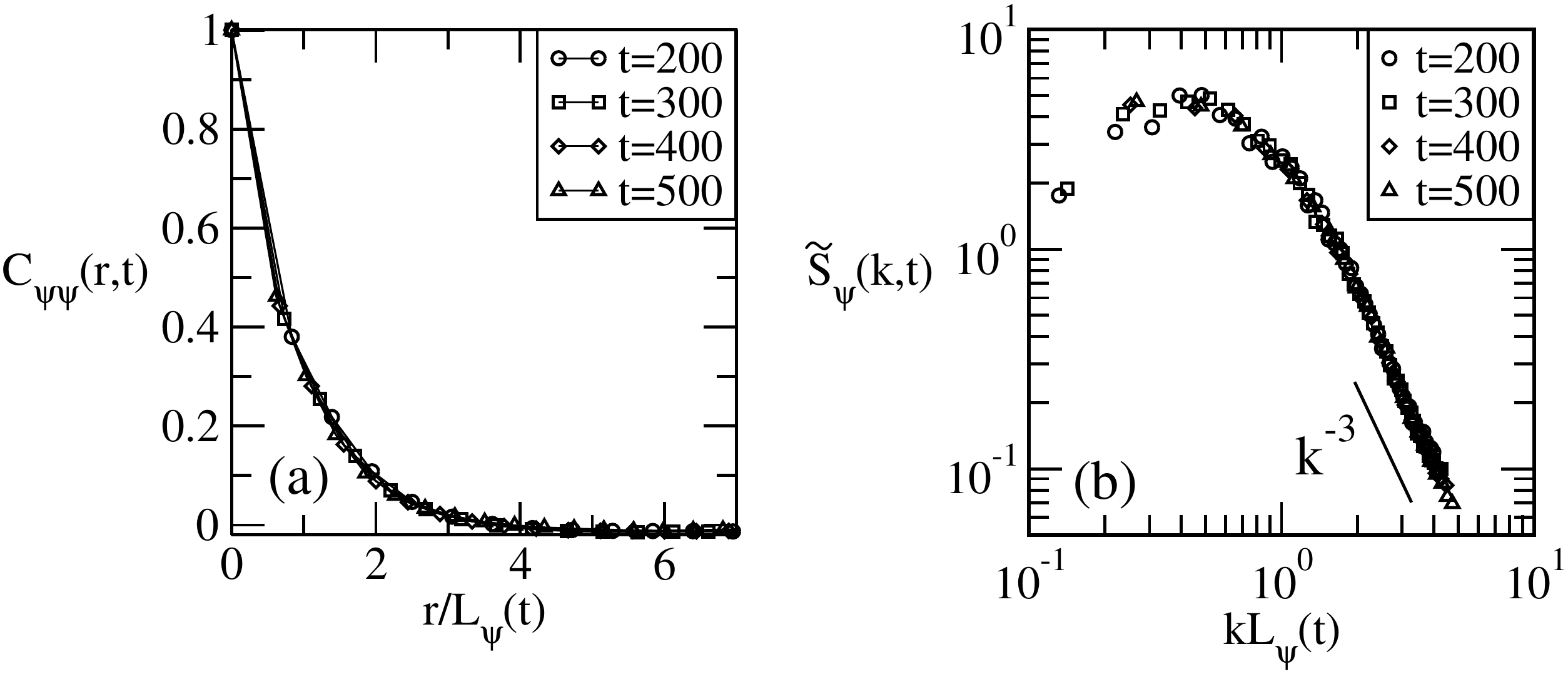}
\caption{\label{fn5} Scaling plot of correlation functions and structure factors for the $\psi(\vec r, t)$-field. We obtain spherically averaged $C_{\psi\psi}(r, t)$ and $\tilde S_{\psi}(k, t)$ as an average over ten independent runs on lattices of size $128^2$, after mapping the actual system of size $912^2$. (a) Plot of $C_{\psi\psi}(r, t)$ $vs.$ $r/L_\psi$ at different times, denoted by the indicated symbols. The data collapse at different time corresponds to dynamical scaling. (b) Plot of $\tilde S_{\psi}(k, t)$ $vs.$ $kL_\psi$ on a \textit{log-log} scale at different times. The solid line labeled with $k^{-3}$ shows the \textit{Porod's law}: 
$\tilde S_{\psi}(k, t)\sim k^{-3}$ in the long wave vector limit.}
\end{figure}

If the clustering is characterized by single length scale $L_\psi(t)$, $C_{\psi\psi}(r, t)$ obeys dynamical 
scaling,
\begin{eqnarray}
 \label{dens}
 C_{\psi\psi}(r, t) = g_\psi\left[\frac{r}{L_\psi(t)}\right],
\end{eqnarray}
where $g_\psi(x)$ is the scaling function and $x$ is the scaling variable. We also compute the \textit{structure 
factor} $S_{\psi\psi}(k, t)$, which is the Fourier transform of $C_{\psi\psi}(r, t)$ at wave vector $\vec k$. The
dynamical scaling form for $S_{\psi\psi}(k, t)$ is given by
\begin{eqnarray}
 \label{denf}
 S_{\psi\psi}(k, t) = L_\psi^d(t)\tilde S_\psi\left[kL_\psi(t)\right],
\end{eqnarray}
where $\tilde S_\psi(p)$ is the scaling function, $d$ is the spatial dimension, and $p$ is the scaling variable. 
All statistical quantities presented here are obtained as averages over ten independent runs. In Fig.~\ref{fn5}, 
we plot equal time correlation functions [$C_{\psi\psi}(r, t)$ in Fig.~\ref{fn5}(a)] and structure factors 
[$S_{\psi\psi}(k, t)$ in Fig.~\ref{fn5}(b)] for the $\psi(\vec r, t)$ field. In Fig.~\ref{fn5}(a), the numerical 
data at different times collapse, confirming dynamical scaling. An important characteristic of the morphology is the \textit{Porod tail}: $\tilde S_{\psi}(k, t)\sim k^{-(d+1)}$ for large $k$~\cite{24}. This is a consequence of scattering from sharp interfaces, which is shown in Fig.~\ref{fn5}(b). However, a deviation from Porod's law may be observed due to the finite thickness or roughness of interfaces~\cite{22}. If $w$ is the thickness of the interface, Porod's law will be observed at large $L_\psi(t)$, i.e., in the limit $w/L_\psi(t)\rightarrow0$.

\begin{figure}
\centering
\includegraphics*[width=0.5\textwidth]{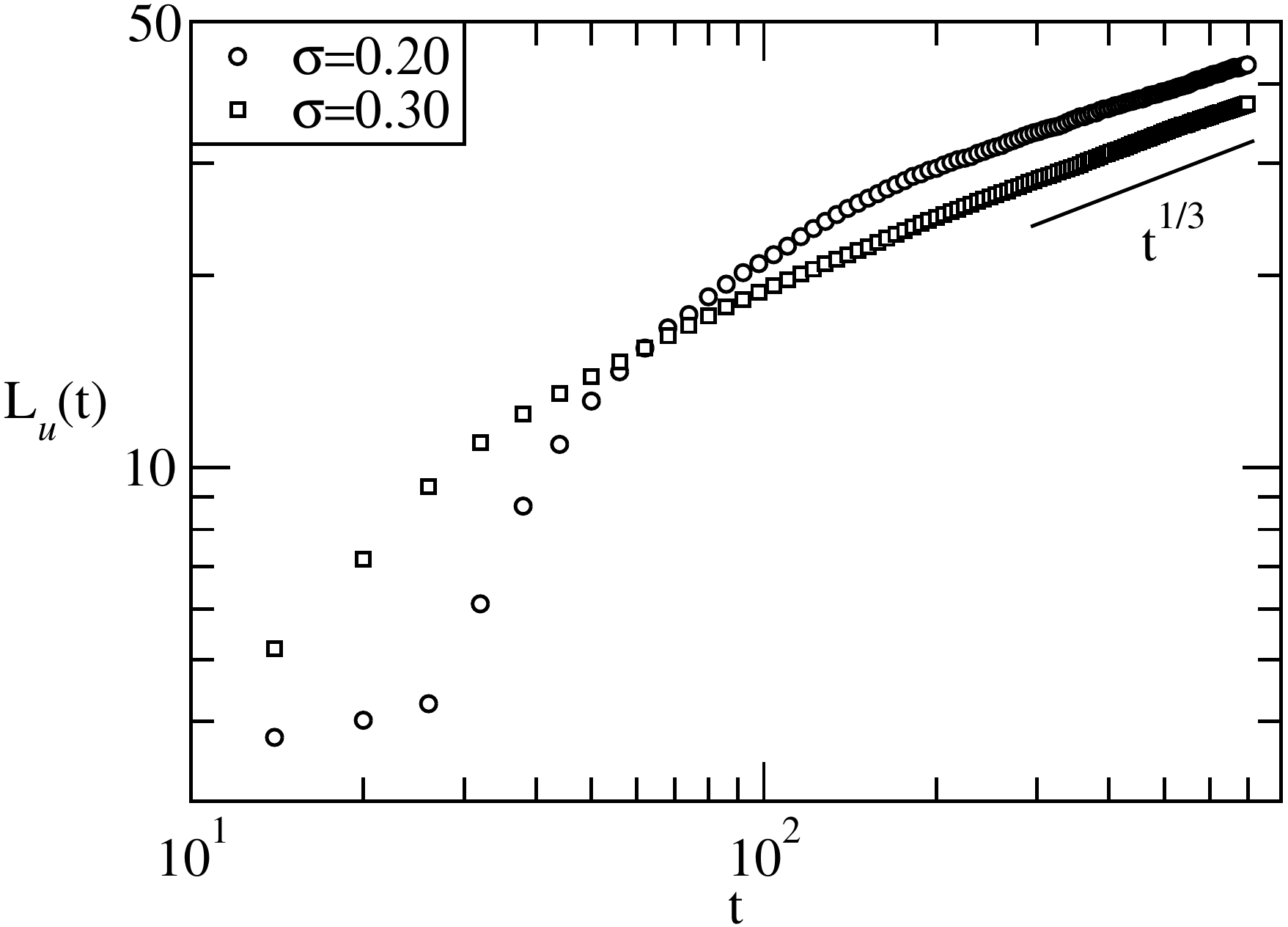}
\caption{\label{fn7} The correlation length scale $L_u(t)$ of the  $\vec u(\vec r, t)$ field as a function of 
time. Plot of $L_u(t)$ $vs.$ $t$ on a \textit{log-log} scale. The solid line with an exponent of $\frac{1}{3}$ is shown, corresponds to diffusive growth for ordering.}
\end{figure}

The evolution snapshots of velocity field shown in Fig.~\ref{fn2} indicate that ordering and pattern formation 
exist also in the velocity field. To clarify the nature of pattern formation~\cite{20,32}, we have hardened the velocity field in Fig.~\ref{fn2}; i.e., the length of all vectors has been set to unity. In fact, this means that we consider instead of the velocity field the direction field $\vec u=\vec v/|\vec v|$, similar to the order-parameter for domain growth in the XY model in two dimensions~\cite{25}. The velocity field is assigned the value zero at points with no particles in the associated coarse-graining box. Void spaces in Fig.~\ref{fn2} correspond to such points. Similar to the $\sigma(\vec r, t)$ field, the evolution of the $\vec v(\vec r, t)$ field is characterized by the velocity correlation function $C_{uu}(r, t)$, defined as
\begin{eqnarray}
 \label{velc}
 C_{uu}(r, t) = \left\langle\vec u\left(\vec R, t\right)\cdot\vec u\left(\vec R + \vec r, t\right)\right\rangle - \left\langle\vec u\left(\vec R, t\right)\right\rangle\cdot\left\langle\vec u\left(\vec R + \vec r, t\right)\right\rangle,
\end{eqnarray}
Similar to the case of the order parameter, we define a correlation length $L_u(t)$ from $C_{uu}(r, t)$, which is shown in Fig.~\ref{fn7}. Clearly, $L_u(t)$ follows diffusive growth: $L_u(t)\sim t^{1/3}$. Details are given in the figure caption.

\begin{figure}
\centering
\includegraphics*[width=0.8\textwidth]{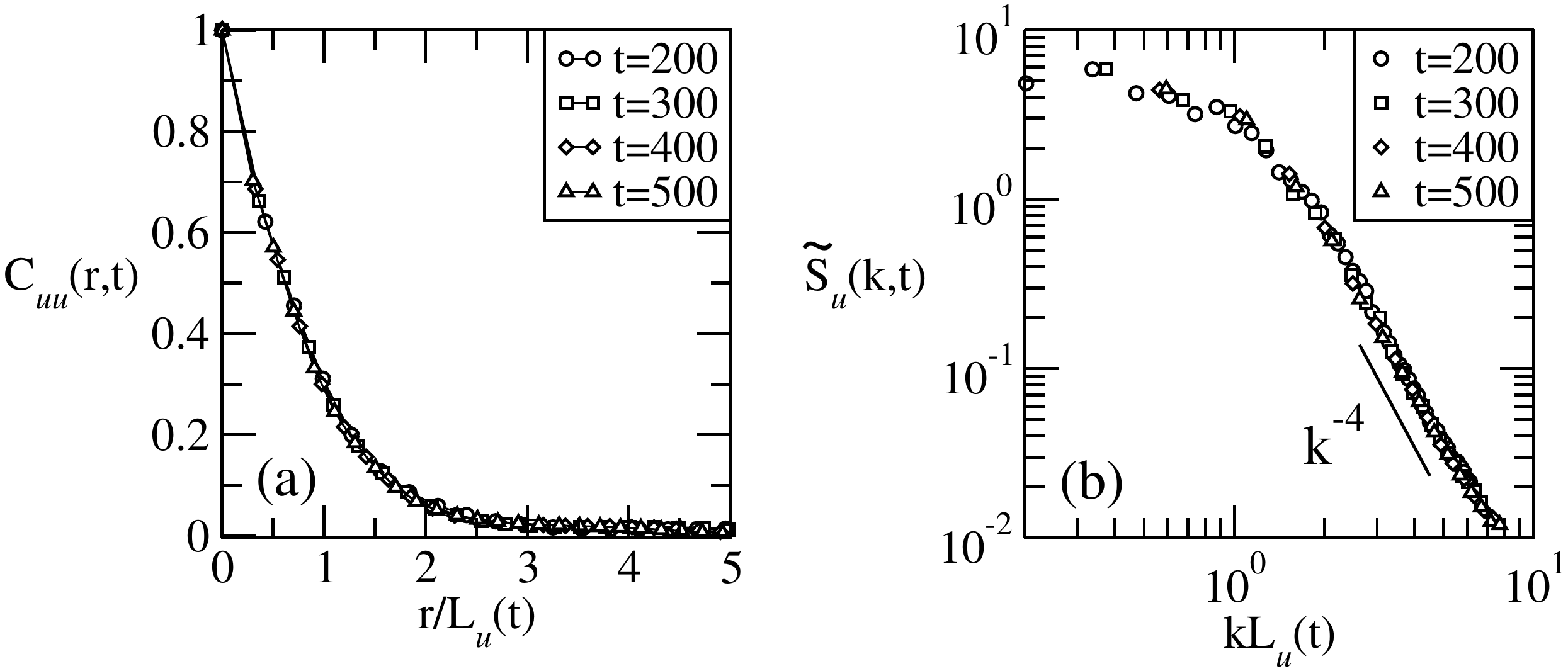}
\caption{\label{fn6} Scaling plot of spherically-averaged correlation functions $C_{uu}(r, t)$ and structure
factors $\tilde S_{u}(k, t)$ for $\vec u(\vec r, t)$-field. (a) Plot of $C_{uu}(r, t)$ $vs.$ $r/L_u$ at different time, denoted by the indicated symbols. The data collapse at different time confirms to dynamic self-similarity. 
(b) Plot of $\tilde S_{u}(k, t)$ $vs.$ $kL_u$ on a \textit{log-log} scale at different times. The solid line labeled with $k^{-4}$ represents the \textit{generalized Porod's law}: $S_{uu}(k, t)\sim k^{-(d+n)}$ for $d=2$ and $n=2$. The rest of the details are same as described in the caption of Fig.~\ref{fn5}.}
\end{figure}

Again the correlation function can be expected to obey dynamical scaling,
\begin{eqnarray}
 \label{vels}
 C_{uu}(r, t) = g_u\left[\frac{r}{L_u(t)}\right],
\end{eqnarray}
where $g_u(y)$ is the scaling function, and $y$ is the scaling variable. The corresponding dynamical scaling form for the \textit{structure factor} $S_{uu}(k, t)$ is given by
\begin{eqnarray}
 \label{velf}
 S_{uu}(k, t) = L_u^d(t)\tilde S_u\left[kL_u(t)\right],
\end{eqnarray}
where $\tilde S_u(q)$ is the scaling function, $d$ is the spatial dimensionality, and $q$ is the scaling variable. In Fig.~\ref{fn6}, we plot equal-time correlation functions [$C_{uu}(r, t)$ in Fig.~\ref{fn6}(a)] and structure factors [$\tilde S_{u}(k, t)$ in Fig.~\ref{fn6}(b)] for the $\vec u(\vec r, t)$ field. Similar to the $\psi(\vec r, t)$ field, the evolution of the $\vec u(\vec r, t)$ field, follows dynamic scaling. In limit $k \rightarrow \infty$~\cite{20,21,22,23}, the structure factor follows \textit{generalized Porod's law} as $S_{uu}(k, t)\sim k^{-(d+n)}$ with $d=2$ and $n=2$, which is the number of components of $\vec u$. This is the consequence of the scattering from vortex-like defects~\cite{25}.

\section{Summary and Discussion}
Let us conclude this paper with summary and discussion of our results. We studied the cooling of low-density 
granular gases using large-scale molecular dynamics, where solid friction between two interacting particles is 
used as the only dissipation mechanism of energy. The system cools down algebraically with a changing exponent.
At the early stage, the exponent describing the cooling down is $\alpha=3/2$ and then the cooling slows down and
is characterized at the late stage by $\alpha=1$. We observe clustering in the density field and local ordering 
in the velocity field because of the cooling, and the velocity distribution, which is originally a
Maxwell-Boltzmann distribution, deviates from it at later times. The morphology of clustering in the density field is studied by obtaining equal-time correlation functions and structure factors, showing dynamical scaling. The average cluster size of the density field shows diffusive growth: $L_\psi(t)\sim t^{1/3}$. Similar to the density field, we also observe dynamical scaling of equal-time correlation functions and structure factors of the velocity field. The characteristic length scale of velocity field shows power law growth: $L_u(t)\sim t^{1/3}$, proposed in the past in the context of dense granular flow~\cite{16}. We hope that the information provided in this paper will be useful for the further study of rheology of dense granular flows.
\vskip 0.2cm
\noindent\textbf{Acknowledgments}\\
\indent P.D. acknowledges financial support from Council of Scientific and Industrial Research, India. S.P. is grateful to UGC, India, for support through an Indo-Israeli joint project. He is also grateful to DST, India, for support through a J. C. Bose fellowship. The research of MS, Grant No. 839/14, was supported by the ISF within the ISF-UGC(Israel) joint research program framework.

\end{document}